\documentclass[11pt,twoside,a4paper]{article}
\usepackage[utf8]{inputenc}
\usepackage{graphicx}
\usepackage[colorlinks,allcolors=cyan!70!black]{hyperref}
\usepackage{amsmath}
\usepackage{multirow}
\usepackage[a4paper, total={6in, 8in}]{geometry}
\usepackage{amsfonts}
\usepackage{subcaption}
\usepackage{hyperref}
\usepackage{float}
\usepackage{xcolor}
\usepackage{authblk}
\definecolor{bordeaux}{RGB}{128,0,32} 
\usepackage{url}
\usepackage{textalpha}
\usepackage{algorithmic}  
\usepackage{graphicx}
\usepackage{booktabs}    
\usepackage{caption}      

\def\IP3{IP\textsubscript{3}}
\def\Ca{Ca$^{2+}$ }
\usepackage{todonotes}

\usepackage{lipsum}

\title{Accounting for Missed Events in the Bayesian Modeling of \IP3R Multimodal Gating
}

\author[1]{Schayma Ben marzougui}
\author[1,2]{Audrey Denizot}
\author[1,3]{Hugues Berry}
\affil[1]{AIstroSight, Inria, Hospices Civils de Lyon, Universite Claude Bernard Lyon 1, F-69603 Villeurbanne, France}
\affil[2]{Correspondence: audrey.denizot@inria.fr}
\affil[3]{Correspondence: hugues.berry@inria.fr}

\begin{document}
\maketitle
\pagestyle{plain}

\begin{abstract}
The Inositol 1,4,5-trisphosphate receptor channel (\IP3R) is an important calcium channel involved in calcium-induced calcium release, playing a prominent role in intracellular calcium signaling. However, accurately characterizing its gating behavior remains a challenge, particularly due to the temporal resolution of patch clamp techniques that is not large enough to detect all short-lived events. This limitation can significantly bias the inference of kinetic models describing the receptor activity.
To address this issue, we focused on the quantitative analysis of \IP3R gating behavior using patch clamp data, with particular attention to missed events. We modeled \IP3R channel gating using Hierarchical Markov chains and used a Bayesian approach that integrates missed event correction directly into the likelihood function, enabling more accurate parameter inference and model evaluation. We show that accounting for missed events deeply clarifies the multi-modal model that emerges from model selection. In this new model, the Park and Drive modes both consist of the same 3-state Markov model, with mode-dependent kinetic parameters: the Drive mode stabilizes the closed state
directly connected to the open one, whereas the Park mode stabilizes the other closed state, that is not connected to the open one. Intermediate \Ca concentrations are found to strongly depress the Drive to Park transition rate, so that the \IP3R channel undergoes frequent transitions to the Park mode only for < 50 nM or micromolar \Ca concentrations. 
Overall, our approach provides a refined perspective on \IP3R channel modeling and highlights the critical importance of accounting for missed events upon model selection based on single-channel recordings.

%Furthermore, we explored model selection using Convolutional Neural Networks \cite{oikonomou_deep_2024}, enabling a wider range of candidate topologies to be evaluated 25 in total compared to the limited set considered in the original study. 

\end{abstract}

\section*{Introduction}
The inositol 1,4,5-trisphosphate receptor (\IP3R) is a ligand-gated \Ca channel located in the membrane of the endoplasmic reticulum. By releasing \Ca from intracellular stores, the \IP3R plays a pivotal role in shaping the spatio-temporal properties of calcium signals such as oscillations and waves~\cite{berridge_calcium_2006}. 

Following the pioneering work of Neher and Sakmann \cite{neher_extracellular_1978, hamill_improved_1981}, the patch-clamp technique has become the reference method for characterizing ion channel kinetics at the single-channel level. Nuclear patch-clamp experiments have provided detailed information on \IP3R gating behavior under varying concentrations of \IP3, \Ca, and ATP \cite{wagner_ii_differential_2012, mak_inositol_2015}. These studies have revealed a modal gating phenomenon, in which the channel switches between low- and high-activity modes depending on the ligand condition. 

Mechanistic descriptions of ion channel dynamics are most commonly based on continuous-time Markov chains (CTMCs), in which the channel is represented as a network of open and closed states connected by stochastic transition rates. For the \IP3R, early models such as the De Young–Keizer model \cite{de_young_single-pool_1992} successfully captured macroscopic \Ca release but did not explicitly incorporate single-channel data. More recent efforts, notably those by Siekmann et al. \cite{siekmann_kinetic_2012}, incorporated patch-clamp recordings within a Bayesian inference framework using Markov chain Monte Carlo (MCMC) sampling to infer Markov models that account for modal gating.
These models provided estimates of kinetic parameters, widely adopted in computational studies of \Ca dynamics such as deterministic models of oscillations \cite{sneyd_dynamical_2017, cao_deterministic_2014} and integro-differential models of \Ca puffs \cite{hawker_ca2_2025}. Furthermore, Siekmann et al. \cite{siekmann_modelling_2016} introduced a hierarchical Markov representation to model \IP3Rs. In this framework, the intra-modal kinetics are described by a set of Markov models $Q_i$ capturing the opening and closing transitions within each mode, while the inter-modal kinetics are governed by a separate Markov process $\tilde{M}$ modeling mode switching. 

Despite the good temporal resolution of patch clamp techniques, a study based on real channel recordings has estimated that these recordings can miss as many as 88\% of short channel closing events \cite{burzomato_single-channel_2004}. Brief openings or closings shorter than the acquisition interval remain undetected, leading to so-called missed events. As emphasized by Colquhoun, Hawkes, and collaborators, neglecting these missed events biases dwell-time distributions and can distort estimated rate constants, particularly those governing fast transitions \cite{colquhoun_joint_1996}. To address this issue, Hawkes and Jalali \cite{hawkes_distributions_1990} derived exact likelihood functions that integrate over undetected events, providing the foundation for maximum-likelihood fitting with missed-event correction, as implemented in computational tools such as \texttt{DCPROGS} \cite{colquhoun_joint_1996}. Building on this foundation, Epstein et al. \cite{epstein_bayesian_2016, ball_asymptotic_2006} introduced a Bayesian framework embedding the corrected likelihood within a MCMC scheme. In contrast, Gin et al. \cite{gin_markov_2009} applied the approximate correction of Blatz and Magleby \cite{blatz_correcting_1986} to adjust open and closed time distributions post-hoc, without incorporating the correction into the model likelihood and without addressing modal gating.\\

The Bayesian approach developed by Siekmann and co-workers \cite{siekmann_kinetic_2012} to infer kinetic parameters of \IP3R gating relies on a discrete-time classification of each sample as open or closed, followed by likelihood computation without explicit treatment of the unobserved channel state between consecutive sampling points. While this formulation avoids the need to reconstruct transition times within each interval, it does not fully overcome the physical limitation imposed by finite sampling. When the acquisition interval is large relative to the kinetics of fast transitions, short-lived ($<$ temporal resolution) events remain invisible to the model. As a result, the likelihood neglects these missed transitions, leading to biased parameter estimates, particularly for fast intra-modal rates, as demonstrated by Epstein et al. \cite{epstein_bayesian_2016}. 

This methodological limitation motivated the present study, in which we reinvestigated \IP3R gating models within a Bayesian framework that integrates exact missed-event correction into intra-modal analysis. This approach enables more reliable parameter inference and refined model topology selection.

%In parallel, recent advances in data-driven modeling and machine learning have opened new perspectives for single-channel analysis. Oikonomou et al. \cite{oikonomou_deep_2024} applied convolutional neural networks (CNNs) to discriminate among large sets of candidate topologies, enabling systematic exploration of model spaces beyond the reach of classical statistical methods. Inspired by this approach, we combine Bayesian inference with exact missed-event correction and CNN-based model selection to expand the range of candidate topologies and improve the robustness of kinetic parameter estimation. Together, these strategies provide a more accurate and comprehensive understanding of IP\textsubscript{3}R gating, emphasizing the critical impact of experimental resolution on quantitative modeling of single-channel dynamics.

\section*{Methods}

\subsection*{Experiments}

We used as raw data the on-nucleus patch-clamp recordings of single \IP3R channel activity reported by Wagner and Yule \cite{wagner_ii_differential_2012}. The data was acquired at 100 mV, sampled at 20 kHz, filtered at 5 kHz, and obtained under controlled \IP3, \Ca, ATP, and BAPTA concentrations. \\

%To analyzed this data, we adopted the Hierarchical Markov models framework for \IP3R modeling developed by Siekmann et al. \cite{siekmann_modelling_2016}, and focused mainly on the intra-mode analysis, which describes the stochastic opening and closing of the channel within modes.\\

\subsection*{Idealization and segmentation}

\begin{figure}[bp!]
    \hspace{-0.2cm}
    \includegraphics[width=\linewidth]{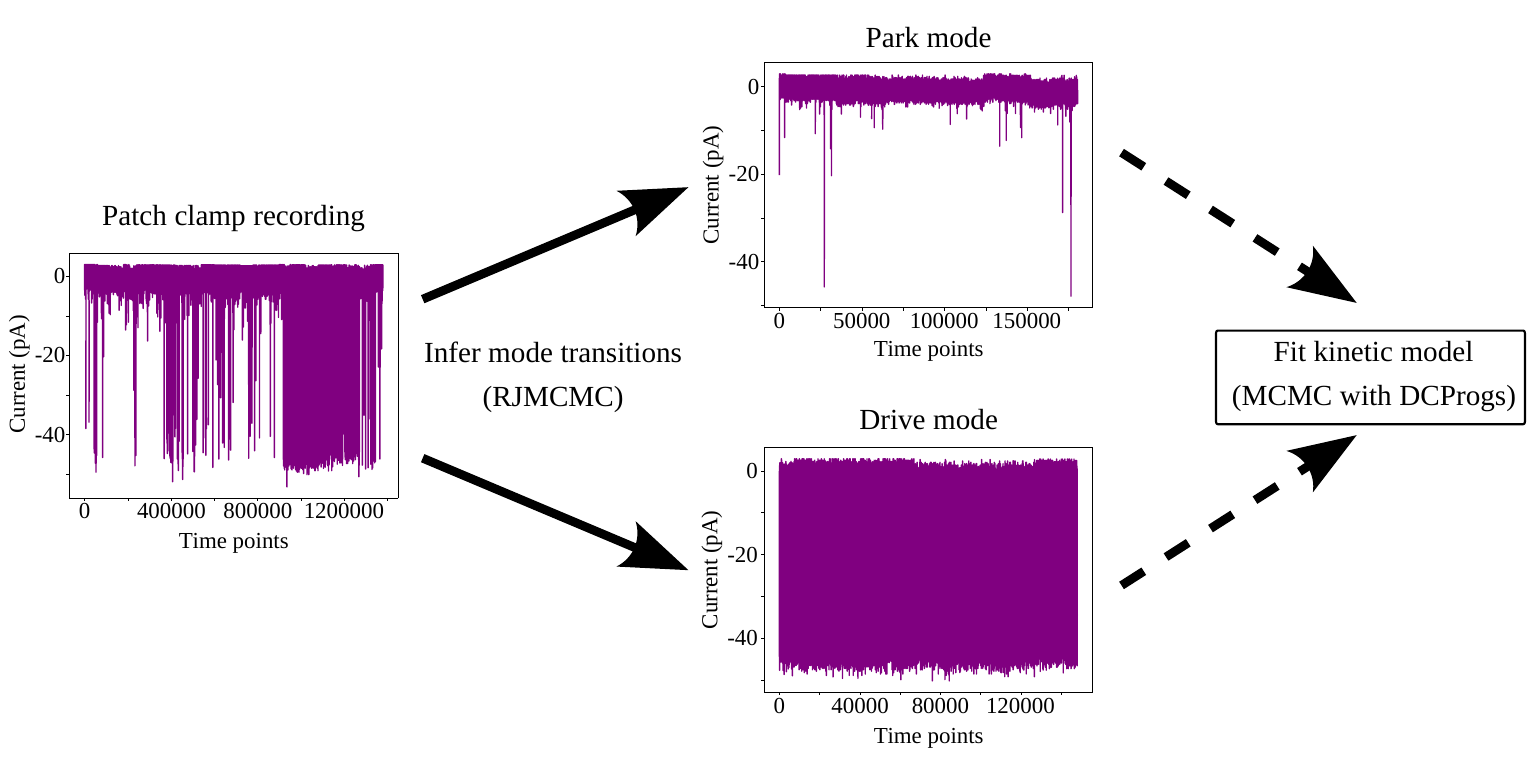}
    \caption{\textbf{Workflow for \IP3R modal gating analysis.} Changepoints between distinct gating modes are identified from patch-clamp traces using Reversible Jump MCMC \cite{siekmann_statistical_2014}. Topologies are further analyzed using MCMC with the DCPROGS framework, including missed-event correction \cite{epstein_bayesian_2016}.}
    \label{fig:workflow}
\end{figure}

Raw current traces were idealized into binary open/closed sequences using a thresholding rule (points above half the mean open current were classified as open). Typically, the current traces alternate between periods of frequent opening and quieter periods, corresponding to distinct gating modes (Fig.~\ref{fig:workflow}). The periods of frequent opening are referred to as the ``Drive'' mode whereas the quiet ones correspond to the ``Park'' mode. Identification of the transitions points between these modes, i.e., inter-mode transitions, was used to isolate the modes, analyze each mode separately, and select a model for each mode (intra-mode analysis). The inter-modal transitions rates were finally estimated as a function of \Ca concentration, yielding the final \IP3R model. 

To locate the inter-mode transitions, we applied the Bayesian segmentation method of Siekmann et al. (2014) \cite{siekmann_statistical_2014}. This technique uses a Reversible Jump Markov Chain Monte Carlo (RJMCMC) algorithm \cite{green_reversible_1995} to infer (i) the number, (ii) the locations of changepoints, and (iii) the mean open probability within each segment. RJMCMC sampling explores models of varying dimensionality through four types of moves, birth, death, shift, and step allowing the chain to adaptively add, remove, or adjust changepoints~\cite{green_reversible_1995}.\\

In this Bayesian framework, the likelihood is computed assuming that, within each segment, the observed sequence of open and closed events follows a binomial distribution, which probability of success equals to the segment mean open probability.\

We consider a sequence $Y=\{y(1),\dots,y(N)\}$ of $N$ data points, where $y(n)=1$ denotes an open event and $y(n)=0$ a closed event.
Let
\[
O(n)=\sum_{k=1}^{n} y(k)
\]
be the number of open events observed up to index $n$. The likelihood is given by : 
\begin{equation}
\mathcal{L}(O \mid j, p) = \prod_{i=0}^{k} p_i^{s_i} (1 - p_i)^{u_i},
\end{equation}

where $k$ is the number of segments detected by the RJMCMC algorithm, $j = \{j_0, j_1, \dots, j_{k-1}\}$ are the changepoints between these segments. $s_i$ and $u_i$ are the number of expected (i.e., undetected) open and closed events, respectively, in segment $i$ and $p_i$ is the mean open probability of segment $i$.\\

Priors are imposed both on the number of changepoints and on their spacing, which prevents unrealistically dense or sparse changepoints. The posterior distribution is then given by:
\begin{equation}
P(j, p \mid O) \propto \mathcal{L}(0 \mid j, p)\, \pi(j,p),
\end{equation}

Note that Siekmann et al.~\cite{siekmann_statistical_2014} also used the recording from Wagner and Yule \cite{wagner_ii_differential_2012} for illustration of their method. We thus could verify as a sanity-check that the number of changepoints we detected in each dataset, as well as the resulting mode-switching frequency, closely matched the values reported in~\cite{siekmann_statistical_2014}. This also allowed us to work with comparable modal segments. 

\subsection*{Characterization of isolated gating modes}

%Following the identification of distinct gating modes using RJMCMC segmentation, we next aimed to determine the most plausible kinetic scheme underlying each mode. 
The dynamics within a single gating mode was modeled as a continuous-time Markov chain (CTMC), in which the channel occupies a finite set of open and closed states and transitions between them according to rate constants. Biophysically, open states correspond to conformations that permit calcium conduction, 
whereas closed states represent non-conducting conformations. \\

In this CTMC approach, the state space \(S_n\) contains \(n_o\) distinct open states and \(n_c\) distinct closed states, so that the total number of states is \(n_s = n_o + n_c\). Transitions between states are encoded in the infinitesimal generator matrix \(Q\), with off-diagonal elements \(q_{ij}\) representing the rate of transition from state \(i\) to \(j\), and diagonal elements defined as
\[
q_{ii} = -\sum_{j \neq i} q_{ij}.
\]

At any time \(t\), the channel occupies a single state \(X_t \in \{1, \ldots, n_s\}\), and waiting times between transitions follow exponential distributions governed by \(Q\).\\

\subsection*{Burst-based analysis}
Intra-modal analysis was then based on the partition of the state sequence into burst series. Following the approach of Colquhoun and Hawkes \cite{colquhoun_joint_1996}, a burst is defined as a sequence of openings and brief closures (shorter than a threshold $t_{\mathrm{crit}}$), separated from other bursts by long closed intervals ($> t_{\mathrm{crit}}$). Each burst begins with an open interval and alternates between open and closed dwell times. \\

In practice, a burst series was constructed as follows:
\begin{enumerate}
    \item Consecutive points in the same state were grouped to form dwell intervals, producing an alternating sequence of open and closed durations.  
    \item The sequence was constrained to start with an open interval and contain an odd number of intervals following \texttt{DCPROGS} requirements. 
    \item The resulting series, representing a single large burst, served as input for the \texttt{DCPROGS} likelihood computation.  
\end{enumerate}

In contrast to classical multi-burst analysis \cite{colquhoun_joint_1996, epstein_bayesian_2016}, our approach focuses on the dynamics within a single gating mode. Following the hierarchical Bayesian framework proposed by Siekmann et al. \cite{siekmann_modelling_2016}, we have made the choice to restrict the analysis to a long, stationary segment of activity representing the intra-modal kinetics of an individual channel. Although on-nucleus patch-clamp recordings may contain several simultaneously active IP\textsubscript{3}R channels \cite{horn_estimating_1991}, our analysis assumes the presence of a single channel. This choice is methodological rather than physiological: it allows us to combine two inference frameworks, one addressing missing event reconstruction and the other describing intra-modal kinetics within a hierarchical model. Consequently, we do not explicitly model multiple channels or apply burst segmentation. Instead, the selected stationary segment is treated as a single continuous burst representing intra-modal dynamics. \\

Once extracted, the burst was interpreted within a CTMC framework, where open intervals correspond to the channel residing in the set of open states \(A\), and closed intervals correspond to the closed states \(F\). The generator matrix is thus partitioned as
$$
Q =
\begin{bmatrix}
Q_{AA} & Q_{AF} \\
Q_{FA} & Q_{FF}
\end{bmatrix},
$$
where $Q_{AA}$ and $Q_{FF}$ describe transitions within open and closed states \cite{colquhoun_stochastic_1982}, respectively, while $Q_{AF}$ and $Q_{FA}$ describe transitions between them. Estimating $Q$ provides a reconstruction of the intra-modal kinetics of the channel.\\

\subsection*{DCPROGS likelihood}

The probability density function (PDF) for observing an open interval of duration $t$, starting in any open state and ending in any closed state \cite{colquhoun_stochastic_1982}, is
$$
f_A(t) = \Phi_A \, G^Q_{AF}(t) \, u_F,
$$
where $\Phi_A$ is the initial distribution over open states, $G^Q_{AF}(t) = \exp(Q_{AA} t) Q_{AF}$, and $u_F$ is a column vector of ones.  

Closed intervals are described analogously by swapping the roles of open and closed states.

\subsubsection*{Correction for missed events}

Because patch-clamp recordings have finite temporal resolution, brief openings or closures 
shorter than the sampling interval \(\tau\) may not be detected, biasing the observed 
dwell-time distributions. To address this, we applied the exact missed-event correction 
originally derived by Hawkes et al. \cite{colquhoun_joint_1996} and implemented in a 
Bayesian context by Epstein et al. \cite{epstein_bayesian_2016}, adapting it to our burst-
based analysis.\\

An apparent open interval of duration \(t\) may contain one or more undetected brief closures (of duration \(< \tau\)). The interval is composed of:
\begin{enumerate}
    \item \textbf{An effective open period of duration \(t - \tau\)}: during which the system resides in \(A\) but may flip to \(F\) and back. This is represented by the survivor function \(R_A(t - \tau)\).  
    \item \textbf{A transition from open to closed states}: described by \(Q_{AF}\), representing the rate of leaving the open states toward the closed states.  
    \item \textbf{A short closure period of duration \(\tau\)}: representing the detected closure that terminates the observed open interval, given by \(\exp(Q_{FF} \tau)\).

\end{enumerate}
Combining these, the missed-event corrected PDF is
$$
G^{Q,e}_{AF}(t) = R_A(t - \tau) \, Q_{AF} \, \exp(Q_{FF}\tau).
$$

For multiple unresolved events within an interval, the full correction involves a convolution of open and closed intervals, handled in the Laplace domain as \cite{colquhoun_joint_1996}:
$$
R_A^*(s) = \left[I - G^{Q,*}_{AF}(s)\, S^*_{FF}(s)\, G^{Q,*}_{FA}(s)\right]^{-1} (sI - Q_{AA})^{-1},
$$
with 
$$
S^*_{FF}(s) = I - \exp[-(sI - Q_{FF})\tau].
$$  
where $G^{Q,*}_{AF}(s)$ denotes the Laplace transform of $G^Q_{AF}(t)$  given by $(sI - Q_{AA})^{-1} Q_{AF}$.\\

The time-domain survivor function, used in the likelihood computation, is obtained via the inverse Laplace transform of \(R_A^*(s)\). This function accounts for brief, undetected closures within open intervals, ensuring that missed events are properly incorporated into the reconstruction of intra-modal kinetics \cite{colquhoun_joint_1996}.\\

Finally, the likelihood of observing a sequence of open and closed dwell times 
$y = (t_1, t_2, \dots, t_m)$
with a temporal resolution of $\tau$ is defined as: 
$$
\mathcal{L}(y\mid Q)
= 
\Phi_A \,
G^{Q,e}_{AF}(t_1)\,
G^{Q,e}_{FA}(t_2)\,
\dots\,
G^{Q,e}_{AF}(t_m)\,
u_F,
$$
where $G^{Q,e}_{AF}(t)$ and $G^{Q,e}_{FA}(t)$ are the missed-event corrected transition matrices,
$\Phi_A$ is the initial distribution over open states, and $u_F$ is a column vector of ones.\\

\subsection*{Bayesian inference and MCMC sampling}

To estimate the rate constants of the infinitesimal generator matrix \( Q \in \mathbb{R}^{n \times n} \) of the Markov model, we employ a Bayesian Markov Chain Monte Carlo (MCMC) approach \cite{hines_primer_2015}. \\

Let \(\theta = (\theta_1, \theta_2, \dots, \theta_d)\) denote the vector of transition rates for a given topology \(S\). The generator matrix \(Q(\theta)\) is constructed according to \(S\) and the current vector of transition rates $\theta$. The likelihood $\mathcal{L}(y \mid Q(\theta))$ is computed from the missed-event formalism described previously.\\

At each MCMC iteration, a new proposal \(\theta'\) is generated independently for each rate using a log-normal distribution:
\[
\theta'_k = \exp\!\bigl[\ln(\theta_k) + \eta_k \bigr], \quad \eta_k \sim \mathcal{N}(0, \, \sigma_k^2),
\]

where $\sigma_k$ is adjusted as follow : 
\begin{itemize}
    \item if the acceptance rate \(< 10\%\): decrease \(\sigma_k \leftarrow 0.9 \, \sigma_k\),
    \item if the acceptance rate \(> 50\%\): increase \(\sigma_k \leftarrow 1.1 \, \sigma_k\).
\end{itemize}

The proposal is accepted with probability
\[
\alpha = \min\!\left( 1, \frac{\mathcal{L}(y \mid Q(\theta')) \, p(\theta')}{\mathcal{L}(y \mid Q(\theta)) \, p(\theta)} \right),
\]

We used weakly informative log-uniform priors over the range \(10^{-5} < \theta_k < 10^5\):
\[
p(\theta) = \prod_{k=1}^d \frac{1}{10^5 - 10^{-5}} \, \mathbf{1}_{\{10^{-5} < \theta_k < 10^5\}}.
\]

After $N$ iterations, the posterior estimate of each rate is taken as the median of the post-burn-in samples. A model selection score is computed using the Bayesian Information Criterion (BIC):
\[
\text{BIC} = -2 \log \mathcal{L}(y\mid Q(\hat{\theta})) + d \log(n),
\]
where $d$ is the number of free parameters and $n$ is the number of observations and the estimator $\hat{\theta}$ is obtained by taking the median of the accepted parameter vectors $\theta$ starting from the burn-in phase.

\subsection*{Software and computational details}

All the analysis code was implemented in Python~3.10 using the \texttt{DCPROGS/HJCFIT} library.
We forked this repository to implement minor adjustments to deal with SWIG errors on modern systems and to facilitate the installation of HJCFIT in a Python virtual environment. The package is available at: \href{https://gitlab.inria.fr/aistrosight/hjcfit}{https://gitlab.inria.fr/aistrosight/hjcfit}.
All code written in support of this publication is available upon request.
The code of the \IP3R model proposed and a simulator is available at \href{https://gitlab.inria.fr/aistrosight/ip3r-model/}{https://gitlab.inria.fr/aistrosight/ip3r-model/}.

\section*{Results}

\begin{figure}[bp!]
    \centering
    \includegraphics[width=\linewidth]{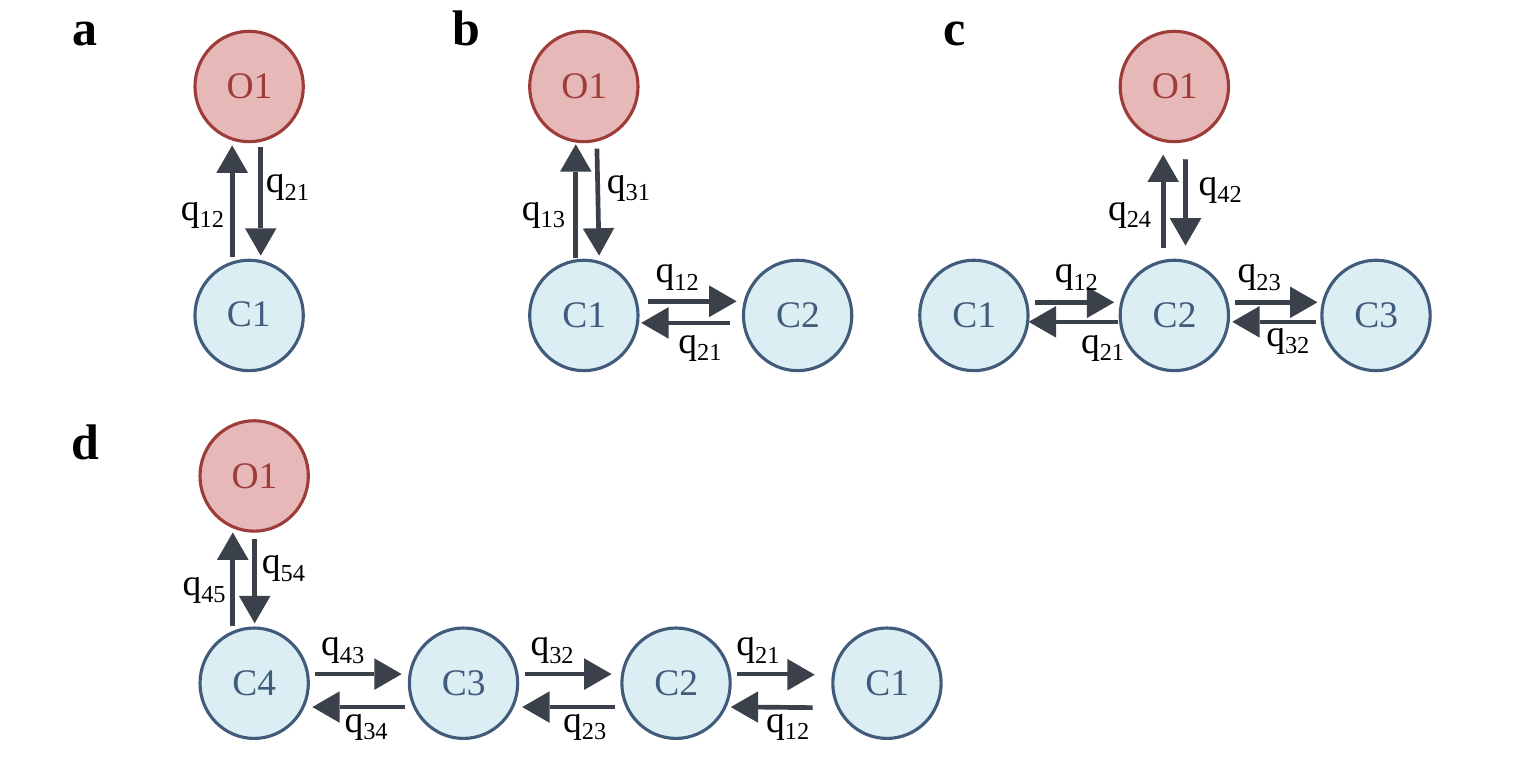} 
    \caption{\textbf{Reference kinetic Markov models for \IP3Rs proposed by Siekmann et al.~\cite{siekmann_kinetic_2012}.} Red circles correspond to open states, and blue ones to closed states. Arrows represent transition rates from one state to another.} 
    \label{fig:1} 
\end{figure}

We applied the modeling framework outlined in Fig.~\ref{fig:workflow}. Single-channel recordings of \IP3R channels are characterized by alternation of periods with frequent transitions to the open state with periods where channel openings are much more infrequent~\cite{Ionescu_2007,siekmann_kinetic_2012}. The former segments are referred to as ``Drive'' mode (frequent openings), and the latter as ``Park'' mode, in analogy with automatic gearboxes (Fig.~\ref{fig:workflow}). First, we used the RJMCMC method of Siekmann et al.~\cite{siekmann_statistical_2014} to identify the sequence of modal states in the IP\(_3\)R2 patch-clamp recordings of Wagner et al.~\cite{wagner_ii_differential_2012}. For each detected mode, we then performed Bayesian parameter inference to determine the most likely topology among a collection of four possible topologies, shown in Fig.~\ref{fig:1}, and to estimate the corresponding intra-modal transition rates.

The set of possible topologies considered in Fig.~\ref{fig:1} originates from a stepwise model building procedure~\cite{siekmann_kinetic_2012}, and was used here as a reference to compare two Bayesian inference methods: a MCMC approach that does not account for missed events~\cite{siekmann_kinetic_2012} and our proposed approach, where the \texttt{DCPROGS} likelihood was applied to account for missed events. This comparison was carried out in two steps: (1) first, we evaluated whether the two methods inferred similar transition rates, topology by topology, and (2) we evaluated whether the two methods selected the same topology as the best one according to the Bayesian Information Criterion (BIC).

 %Starting from the simplest posiible two-state open–closed scheme, they incrementally added open or closed states and assessed whether each extension produced an improvement in the likelihood. Models for which the added states failed to increase the likelihood, or for which the newly introduced states exhibited negligible stationary probability, were not selected.

Below, we start by comparing the inferred transition rates and model selection using the inference results averaged across the \IP3R2 datasets recorded at different \Ca concentrations (10, 50, 200,  1000, 5000, and 10000~nM, with [\IP3]=10 $\mu$ M and [ATP] = 5~mM ATP). Given the importance of \Ca regulation for the \IP3R channel, we then provide a more detailed analysis, performed independently for each \Ca concentration.

\subsection*{Intra-Modal kinetics inferences averaged over all Ca\(^{2+}\) concentrations}

\subsubsection*{Comparison of the inferred transition rates between the MCMC \& \texttt{DCPROGS} approaches}

\begin{figure*}[t!]
    \includegraphics[width=0.9\linewidth]{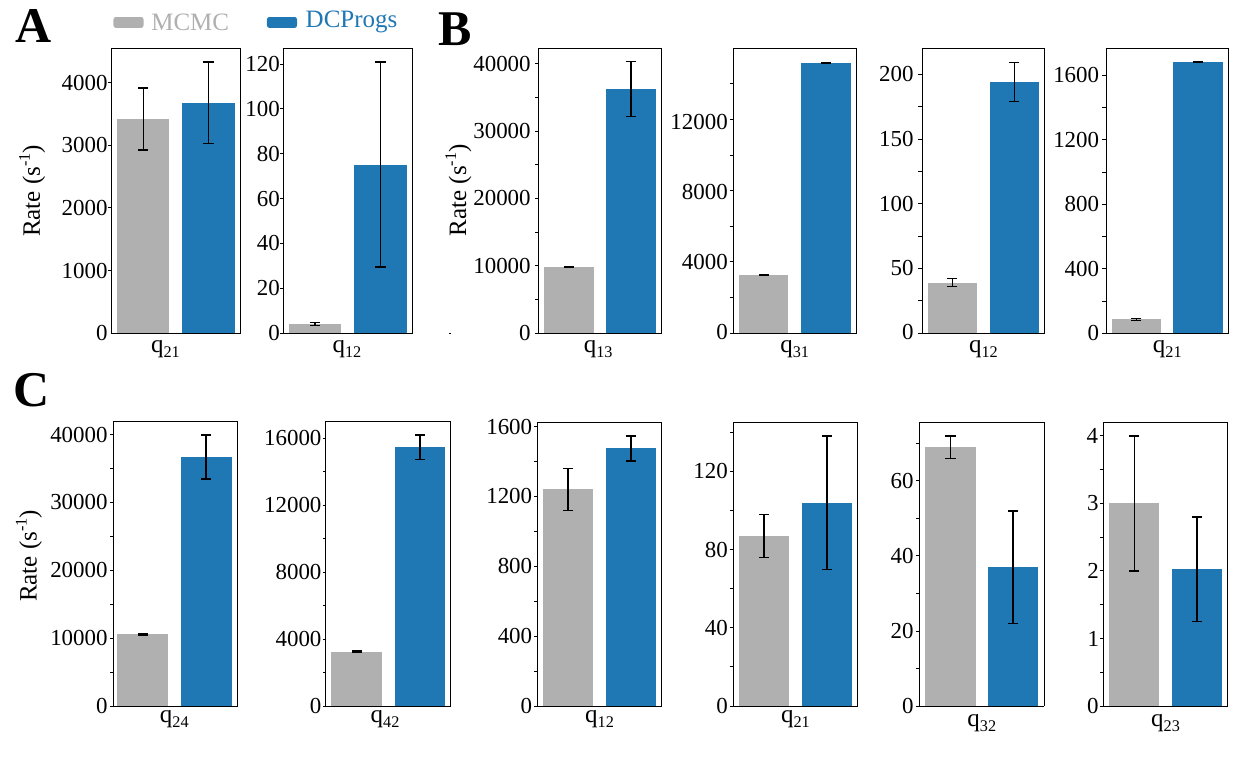}
    \centering
    \caption{\textbf{Comparison of transition rates inferred from \IP3R2 patch-clamp recordings using the classical MCMC method and our \texttt{DCPROGS}-based inference.} Each box plot shows the transition rates estimated across all \IP3R2 datasets at 10~µM \IP3, 5~mM ATP, and various \Ca concentration (10, 50, 200, 1000, 5000, and 10000~nM) from Wagner \& Yule~\cite{wagner_ii_differential_2012}.
    Transition rate inference is shown for the Park mode in topology (a) (Fig~\ref{fig:1}a, A), and for the Drive mode in topologies (b) (Fig \ref{fig:1}b, B), and (c) (Fig~\ref{fig:1}c, C). Transition rates are depicted in gray and blue when inferred using the classical MCMC or our \texttt{DCPROGS}-based inference method, respectively.
    %The left panel shows the rate pair $(q_{62}, q_{26})$, for which the difference between the two inference methods is particularly marked.  
    %The right panel shows the remaining rates $(q_{12}, q_{21}, q_{32}, q_{23})$, for which both inference approaches give similar results. 
    Error bars indicate posterior standard deviations.}
    \label{fig:3}
\end{figure*}

Figure~\ref{fig:3}A shows the transition rates inferred for the simplest topology possible, with only one open and one closed state (Fig.~\ref{fig:1}a) on Park mode, i.e. the mode where opening transitions are less frequent, as topology (a) is selected by the classical MCMC approach for the Park mode. Both inference methods produced similar estimates for the transition rate $q_{21}$, but markedly diverged for $q_{12}$.
Indeed, the classical MCMC method produced mean rates of $\langle q_{21} \rangle = 3420\,\text{s}^{-1}$ and $\langle q_{12} \rangle = 4.14\,\text{s}^{-1}$, whereas our \texttt{DCPROGS}-based approach yielded $\langle q_{21} \rangle = 3681\,\text{s}^{-1}$ and a markedly higher mean $\langle q_{12} \rangle = 75\,\text{s}^{-1}$.  
This difference reflects the detection of short-lived openings that the likelihood used in the classical MCMC method did not detect. 

\renewcommand{\thesubfigure}{\roman{subfigure}}

\begin{figure*}[th!]
    \centering
    \includegraphics[width=\linewidth]{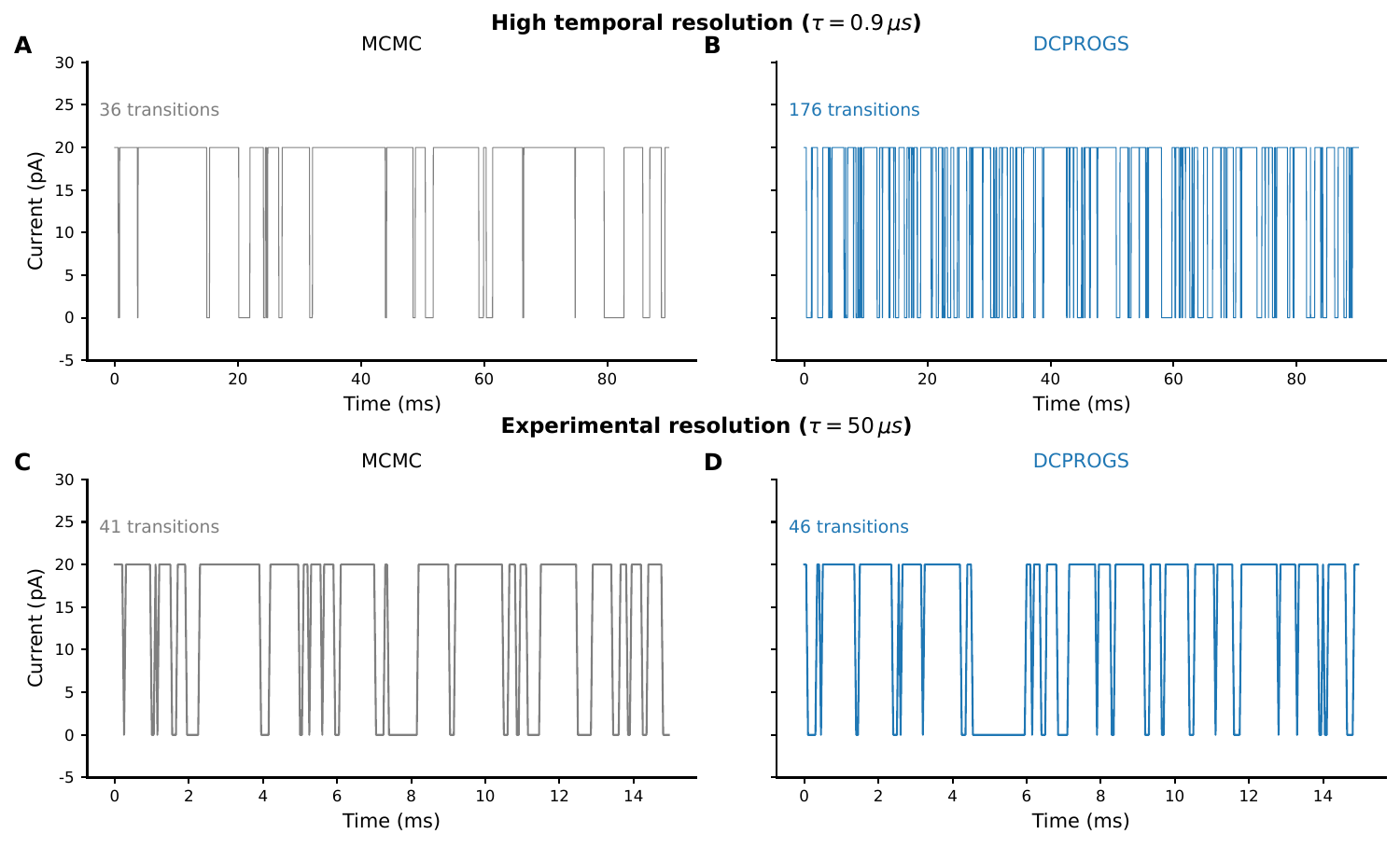}
    \caption{
    \textbf{Comparison of simulated single-channel \IP3R currents obtained using transition rates inferred with the classical MCMC approach or our \texttt{DCPROGS}-based method.}
    Simulated traces are shown for topology (b) (Fig~\ref{fig:1}) in Drive Mode at high temporal resolution ($\tau = 0.9~\mu$s; A,B) and at experimental resolution ($\tau = 50~\mu$s; C,D). \IP3R currents inferred with the classical MCMC approach are depicted in grey (panels A,C) and with our \texttt{DCPROGS}-based method in blue (panels B,D).}
    \label{fig:ip3r_comparison}
\end{figure*}

We then compared the inference results for the Drive mode and topology (b) (Fig~\ref{fig:3}B). Here again, the transition rates estimated with \texttt{DCPROGS} were significantly faster than those estimated using the classical MCMC method \cite{siekmann_kinetic_2012}, especially for transitions between the open and closed states. The mean rate $\langle q_{13} \rangle$, which predicts the opening of the system, is three times higher with \texttt{DCPROGS}, and the closed to closed $q_{12}$ rate demonstrates an even greater disparity where $\langle q_{12} \rangle = 39\,\text{s}^{-1}$ for the classical MCMC method, whereas our \texttt{DCPROGS}-based approach yielded $\langle q_{21} \rangle = 194\,\text{s}^{-1}$. This shows that our \texttt{DCPROGS}-based method captures many brief events that are not detected with the classical MCMC approach. To provide a concrete illustration of what this means, we show in Fig.~\ref{fig:ip3r_comparison} two simulation results generated with a time resolution much larger than available in patch-clamp recordings: panel A shows a simulated current using the parameters inferred with the classical MCMC approach whereas panel B provides a simulation with the parameters inferred with our \texttt{DCPROGS}-based method (using the values of Drive Mode and topology (b) above). Clearly, simple visual inspection shows that the single-channel \IP3R currents simulated using the transition rates inferred using our \texttt{DCPROGS}-based method display much more frequent state transitions than with the classical MCMC method (176 vs. 36 transitions, respectively, at high temporal resolution over an 80 ms interval).

Finally, we show in Fig~\ref{fig:3}C the inference results for the Drive Mode and topology (c). The two inference methods yield consistent results for the slower transition rates, but discrepancies appear in the faster ones. This is particularly striking for the pair $(q_{24}, q_{42})$: the classical MCMC method reported mean rates $(\langle q_{24} \rangle,\, \langle q_{42} \rangle) = (10600,\, 3270)$ $s^{-1}$, where our \texttt{DCPROGS}-based approach inferred mean rates three- to five-fold faster $(36716,\, 15478)$ $s^{-1}$.

\subsubsection*{Model selection differences between the MCMC \& \texttt{DCPROGS} approaches}

We next applied the Bayesian Information Criterion (BIC) to the reference topologies of Fig.~\ref{fig:1} to determine which topology was selected as the best depending on the inference scheme. As above, we computed the BIC across multiple \IP3R2 datasets covering \Ca concentrations ranging from 10 nM to 10 µM (5~mM [ATP], 10~µM [IP$_3$]). To decide between two models, we used the usual rule of thumb \cite{kass_bayes_1995}, according to which a BIC difference between 2 and 6 indicates moderate evidence  (strong evidence for more than 6).

The results are shown in table~\ref{tab:bic_comparison} for a \Ca concentration of 10 nM. Our \texttt{DCPROGS}-based approach selected model (b), both for Park and Drive modes. The selection results for the other \Ca concentrations are provided in Table S1. Note that for [\Ca] = 200 nM, both the number and duration of the Park segments were too small to allow for reliable inference. Strikingly, we obtained the same results for almost all available \Ca concentrations: topology (b) was selected in almost all cases. The only two exceptions were observed for the largest \Ca concentrations: our \texttt{DCPROGS}-based approach selected model~(c) for Drive at 5~$\mu$M [\Ca] and for Drive at 10~$\mu$M [\Ca] model~(b) was selected, although similar to model~(c).  In all other tested concentrations, model (b) was preferred. We therefore conclude that topology~(b) is the best model for both Drive and Park modes according to our \texttt{DCPROGS}-based approach.\\
This result is in strong contrast with the results obtained using the classical MCMC method, as reported in \cite{siekmann_kinetic_2012}, where the best model was topology (a) for Drive and topology (c) for Park. This result shows that the improved handling of missed events in \texttt{DCPROGS} influences not only parameter estimation but also the choice of the best model topology.

\begin{table}[H]
\centering
%\Large
\begin{tabular}{lccccc}
\hline
 & \multicolumn{4}{c}{\textbf{BIC for Model}}&  \\
\textbf{Mode} & \textbf{a} & \textbf{b} & \textbf{c} & \textbf{d}  & \textbf{Best Model}\\
\hline
Park  & $-559.61$   & $\mathbf{-741.04}$ & $-735.99$ & $-730.30$  & Model b\\
Drive & $-5005.98$  & $\mathbf{-5228.95}$ & $-5220.79$ & $-5216.45$ & Model b\\
\hline
\end{tabular}%
\caption{
\textbf{Comparison of Bayesian model selection for intra-modal \IP3R2 gating models when missed events are accounted for with our \texttt{DCPROGS}-based approach.}
The Bayesian Information Criteria (BIC) evaluated from experimental single-channel recordings obtained at 10~µM IP$_3$, 5~mM ATP, and 10~nM \Ca~\cite{wagner_ii_differential_2012} using our \texttt{DCPROGS}-based approach are displayed for both Park and Drive modes. The models were selected among the topologies (a), (b), (c), and (d) presented in Fig.~\ref{fig:1}.
Lower BIC values indicate better model fits after penalizing for model complexity.
}
\label{tab:bic_comparison}
\end{table}

\subsection*{\Ca-dependent intra-Modal analysis}

We next provide a detailed analysis of the dependence of our inference results on \Ca concentration. To that aim, we fixed below the intra-modal topology to model (b), which was identified as the most plausible model according to the BIC criterion above and systematically applied our \texttt{DCPROGS}-based approach to infer the kinetic parameters of the model.

\subsubsection*{Inference results for the Park Mode}
\begin{figure*}[th!]
    \centering
    \includegraphics[width=\linewidth]{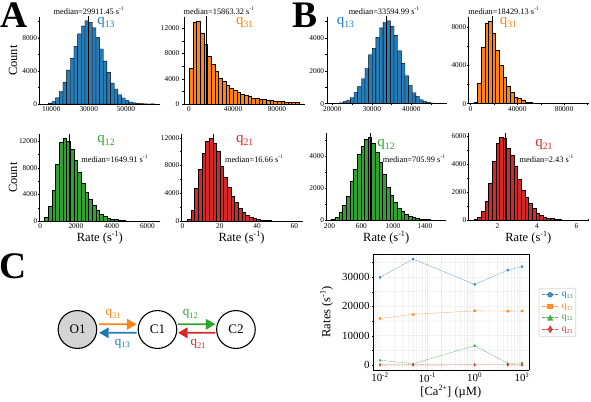}
    \caption{
        \textbf{Posterior distributions and dependence on \Ca concentration of the inferred rates for the Park Mode.}
        Posterior distributions of intra-modal rate constants using model (b) inferred using our \texttt{DCPROGS}-based approach are displayed at two representative calcium concentrations: 0.01 $\mu$M (A) and 10 $\mu$M (B). Full vertical lines locate median posterior values. (C) (Left) Model selected for the Park mode, model (b), with one open state (O1) and two closed states (C1 and C2). (Right) Variation of the intra-Park mode transition rates across the available \Ca concentration range, with [\IP3] = 10 $\mu M$, and [ATP] = 5 mM.
    }
    \label{fig:M1_results_posteriors}
\end{figure*}

The full posterior distributions for all the inferred parameters of model (b) at two representative \Ca concentrations, $0.01\,\mu$M and $10\,\mu$M, are shown in Fig.~\ref{fig:M1_results_posteriors}A and ~\ref{fig:M1_results_posteriors}B, respectively. These distributions show unimodal shapes with well-defined peaks at both $0.01\,\mu$M and $10\,\mu$M $Ca^{2+}$, confirming the quality of the MCMC convergence. While the posterior distributions for the transitions from and to the open state ($q_{31}$ and $q_{13}$) are similar for the two \Ca concentrations, the transitions between the two closed-states seem to be affected by the change of \Ca concentration: the largest \Ca concentration appears to reduce both $q_{12}$ and $q_{21}$ compared to the smallest one.\\
However, examination of the evolution of the inferred rates over the whole range of \Ca concentration (Fig.~\ref{fig:M1_results_posteriors}C) suggests that the four parameters remain essentially \Ca-independent, despite fluctuations that are probably reflecting the error inherent to our inference method and that also explain the apparent change of $q_{12}$ and $q_{21}$ above. Indeed, the Park mode corresponds to a state with infrequent opening events, for which the inference is not as accurate as the active Drive mode and its much larger frequency of opening events. The \Ca-dependence of $q_{13}$ and $q_{12}$ might be anti-correlated (Fig.~\ref{fig:M1_results_posteriors}C), which may indicate a difficulty to infer these parameters independently. Note that for $0.02\,\mu$M [\Ca], both the number and the duration of the Park segments were too small to allow reliable inference. Therefore, these conditions are not shown in the figure. We conclude that the Park mode kinetics parameter inferred with our \texttt{DCPROGS}-based approach are essentially \Ca-independent, which is in agreement with the analysis of Siekmann et al. \cite{siekmann_mcmc_2012}, who used a classical MCMC method without accounting for missed events.

\subsubsection*{Inference results for the Drive Mode}

\begin{figure*}[!t]
    \centering
        \includegraphics[width=\linewidth]{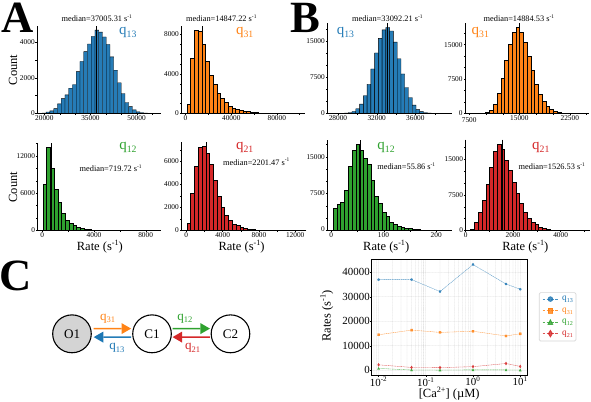}
    \caption{
        \textbf{Posterior distributions and dependence on \Ca concentration of the inferred rates for the Drive Mode.}
        Posterior distributions of intra-modal rate constants using model (b) inferred using our \texttt{DCPROGS}-based approach are displayed at two representative calcium concentrations: 0.01 $\mu$M (A) and 10 $\mu$M (B). Full vertical lines locate median posterior values. (C) (Left) Model selected for the Drive mode, model (b), with one open state (O1) and two closed states (C1 and C2). (Right) Variation of the transition rates across the available \Ca concentration range, with [\IP3] =10 $\mu M$, and [ATP] = 5 mM.
    }
    \label{fig:M2_results_posteriors}
\end{figure*}

Bayesian inference was conducted on Drive mode segments extracted from the same patch-clamp datasets that were used for the Park mode above \cite{wagner_ii_differential_2012} and with the same MCMC parameters, e.g. 100\,000 iterations with a burn-in of 50\,000.
%under identical recording conditions (10~µM IP$_3$, 5~mM ATP, varying Ca$^{2+}$ concentrations). 
Inference was also performed using model topology (b) with our \texttt{DCPROGS}-based approach to infer the kinetics parameters.
%, which was identified as the most plausible model for M1 according to the BIC criterion. 
%The same MCMC conditions as for Mode M2 were applied (100\,000 iterations with a burn-in of 50\,000).
The posterior distributions obtained at \Ca concentrations of 0.01\ and 10 $\mu$M [Ca$^{2+}$], are shown in Fig.~\ref{fig:M2_results_posteriors}A and ~\ref{fig:M2_results_posteriors}B, respectively. Like for the Park mode above, these histograms display unimodal shapes with well-defined peaks. Here again, posterior distributions for the transitions from and to the open state ($q_{31}$ and $q_{13}$) do not vary much between the \Ca concentrations, whereas those between the two closed-states ($q_{12}$ and $q_{21}$) seem to decrease at the largest \Ca concentration. The evolution of the inferred transition rates for all \Ca concentrations is summarized in Fig.~\ref{fig:M2_results_posteriors}C. \\
However, as in the Park Mode, we conclude that the rates display only minor variation over the tested \Ca range, suggesting that intra-modal kinetics in the Drive Mode are also \Ca-independent.

%\subsubsection*{Mode-Dependent Stability and Inversion of Inferred Transition Rates}

\begin{figure}[h]
    \centering
    \includegraphics[width=0.7\linewidth]{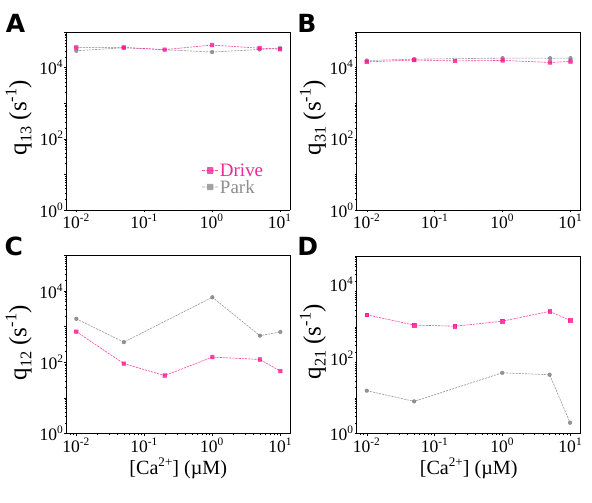}
    \caption{
    \textbf{Comparison of intra-modal transition rates in the Drive and Park modes as a function of \Ca concentration.} 
    %(\textbf{a}) $q_{13}$, (\textbf{b}) $q_{31}$, (\textbf{c}) $q_{12}$, and (\textbf{d}) $q_{21}$.  
    This figure is a replot of Fig.~\ref{fig:M1_results_posteriors}C and Fig.~\ref{fig:M2_results_posteriors}C, providing a direct comparison of the impact of the modes on the inferred values of parameters  $q_{13}$ (A), $q_{31}$ (B), $q_{12}$ (C), and $q_{21}$ (D). Transition rates inferred from topology (b) (Fig~\ref{fig:1}b) are depicted in pink and gray, for the Drive and Park modes, respectively.}
    \label{fig:mode_comparison}
\end{figure}

To better compare the impact of the modes on the \Ca dependence of the parameters, we replotted the results of Fig.~\ref{fig:M1_results_posteriors}C and Fig.~\ref{fig:M2_results_posteriors}C on a parameter-by-parameter basis in Fig.~\ref{fig:mode_comparison}. This figure evidences that the transition rates between the close and open states, \(q_{13}\) and \(q_{31}\) are not only largely \Ca-independent, but also mode-independent, i.e. these rates are hardly sensitive to the choice of mode. In contrast, the figure shows that \(q_{12}\) and \(q_{21}\) exhibit an inverted behavior depending on the mode that explains the overall mode activity: \(q_{12}\) is much larger in Park than in Drive mode, while, on the contrary, \(q_{21}\) is larger in Drive than in Park mode. This suggests a simple explanation of the difference in activity between the two modes, i.e.  segments of frequent opening vs quieter segments in the Drive and Park modes, respectively. Indeed, it seems that the difference of activity is not due to differences in the transitions between open and close states, but is attributable to differences in the transitions between the two closed states. The dwell time in the closed state that is furthest from the open state (C2 in model b) is larger in Park that Drive. In other words, Park corresponds to a stabilization of the closed state that is furthest from the open state (C2). As a result, the \IP3R spends less time in the closed state C1 in Park than in Drive mode. This reduces the possibility to transition to the open state, thus explaining why the opening frequency is severely reduced in Park compared to Drive mode.\\   

Therefore, the suggestion of our \texttt{DCPROGS}-based analysis that the intra-mode model topology is the same in both the Drive and Park modes strongly simplifies the interpretation of the overall dynamics: Drive is found to stabilize the closed state that is directly connected to the open state, thus allowing frequent openings, whereas Park stabilizes the closed state that is not directly connected to the open one, resulting in less frequent openings.

\subsection*{Adding inter-mode transitions: a new model for \IP3R dynamics}
\subsubsection*{Inference of the inter-mode kinetic parameters}
\begin{figure}[H]
    \centering
    \includegraphics[width=0.7\linewidth]{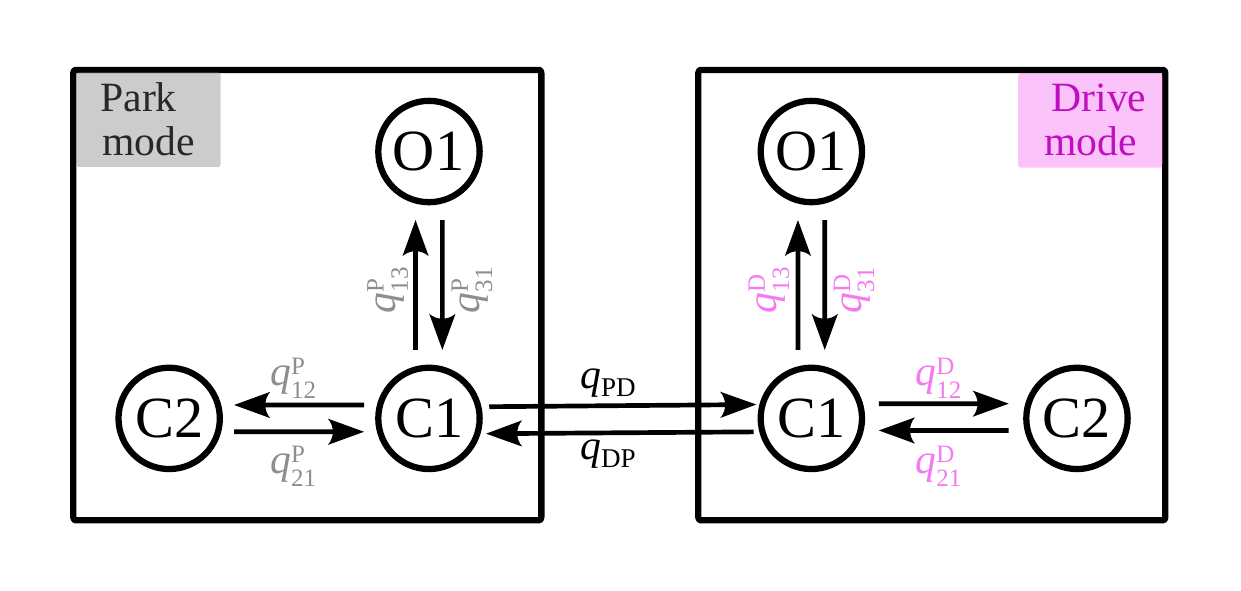}
    \caption{
    \textbf{An \IP3R2 Markov model that takes into account missed events.}
    The \IP3R channel can be in two modes, Park and Drive, that share the same topology but differ by the values of the intra-modal transition rates $q_{ij}^{\text{P}}$ and $q_{ij}^{\text{D}}$. The intra-modal transition rates were inferred using our \texttt{DCPROGS}-based approach. The resulting values are presented in Table~\ref{tab:siekmann_parameters} assuming that they are not \Ca-dependent, as suggested by our results. Inter-modal transition rates
    %Average DCPROGS estimates for the M1 (Park) and M2 (Drive) modes across all Ca\(^{2+}\) concentrations, connected by the inter-modal transition rates 
    \(q_{\text{PD}}\) and \(q_{\text{DP}}\) are hypothesized to connect the two C1 states in each mode. Their \Ca dependence is shown in Fig.~\ref{fig:dcprogs_inter_mode_single}. 
    }
    \label{fig:siekmann_comp_single}
\end{figure}

To describe the transitions between Park and Drive modes, we follow the original model of Siekmann et al.\cite{siekmann_kinetic_2012} and assume that the transitions between the two modes are implemented by a single pair of transitions between the two closed states that are directly connected to their respective open states, i.e. inter-mode transitions are assumed to be transitions between the C1 states of each mode (Fig.~\ref{fig:siekmann_comp_single}).\\

\begin{figure}[tbh!]
    \centering
    \includegraphics[width=0.7\linewidth]{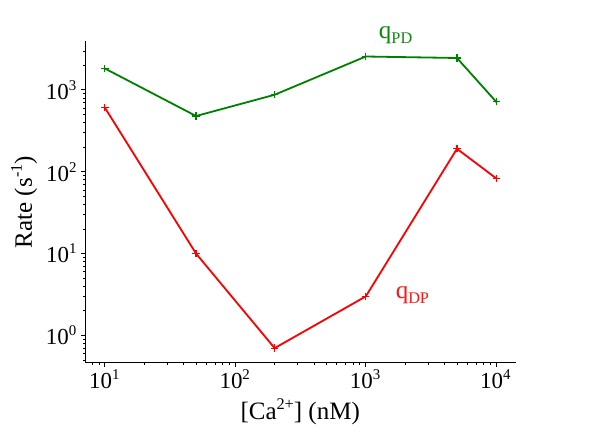}

    \caption{
    \textbf{Inter-modal transition rates estimated with \texttt{DCPROGS}.}
    Inferred inter-modal transition rates \(q_{\text{PD}}\) (Park $\to$ Drive) and \(q_{\text{DP}}\) (Drive $\to$ Park) for \IP3R2 at different \Ca concentrations, 10 µM \IP3, 5 mM ATP. Both Park and Drive modes were modelled using the intra-modal topology of Fig.~\ref{fig:siekmann_comp_single} with the intra-modal kinetic parameters shown in table~\ref{tab:siekmann_parameters}. 
    }
    \label{fig:dcprogs_inter_mode_single}
\end{figure}

In the previous sections, our \texttt{DCPROGS}-based approach suggested that the intra-modal parameters $q_{12}$, $q_{21}$, $q_{13}$, and $q_{31}$ can be mode-dependent but exhibit essentially no dependence on \Ca concentration. We therefore average their inferred mean transition rates across all tested \Ca concentrations. The resulting parameter values are shown in table~\ref{tab:siekmann_parameters}, using $q_{ij}^m$ to refer to the value of the transition rate between state $i$ and $j$ in mode $m=$\{P,D\} for (P)ark and (D)rive, respectively.\\
With the intra-modal parameters thus fixed, we then used our \texttt{DCPROGS}-based approach to infer the values of the inter-mode transition rates, \(q_{DP}\) and \(q_{PD}\). As can be seen from Fig.~\ref{fig:dcprogs_inter_mode_single}, our analysis shows that the inter-modal transition rates have a contrasted dependence on \Ca concentration. The kinetic rate for the transition from the Park to the Drive mode, \(q_{PD}\), was found essentially \Ca-independent. In contrast, the rate for the transition from Drive to Park, \(q_{DP}\), exhibited a major collapse for \Ca concentrations in the [50,1000] nM range, with a decay spanning three orders of magnitude between 10 and 200 nM (607 and 0.69 s$^{-1}$, respectively). This means that intermediate \Ca concentrations strongly depress and can even almost switch off the inter-modal transition to the Park mode. As a result, the \IP3R channel stays almost exclusively in the Drive mode for [\Ca]$\in$[50,1000] nM.\\ 

\begin{table}[tbh!]
\centering
\begin{tabular}{lcc}
\hline 
\textbf{Parameter (s$^{-1}$)} &
\textbf{Park ($m$=P)}&
\textbf{Drive ($m$=D)} \\
\hline
$q_{13}^m$ & 31927 $\pm$ 2984  & 36279 $\pm$ 3585\\
$q_{31}^m$ & 17692 $\pm$ 1016& 15186 $\pm$ 825\\
$q_{21}^m$ & 24 $\pm$ 19      & 1682 $\pm$ 602\\
$q_{12}^m$ & 1979 $\pm$ 2364 & 194 $\pm$ 49\\
\hline
\end{tabular}
\caption{\textbf{Intra-modal parameter values for the Park and Drive modes.} Park ($m$=P) and Drive ($m$=D) modes intra-modal kinetic rates were estimatesd from \IP3R2 single-channel recordings at 10~µM \IP3 and 5~mM ATP. Values are averages across different \Ca concentrations $\pm$ standard deviation. $q_{ij}^m$ refers to the inferred value of the transition rate from state $i$ to $j$ in mode $m=\{\text{P,D}\}$.}
\label{tab:siekmann_parameters}
\end{table}

\begin{table}[tbh!]
\centering
\begin{tabular}{lcc}
\hline 
\textbf{[Ca\(^{2+}\)] (µM)} &
\textbf{$q_{\text{PD}}$ ($\mathrm{s^{-1}}$})&
\textbf{$q_{\text{DP}}$ ($\mathrm{s^{-1}}$}) \\
\hline
$0.01$ & 1829 $\pm$ 206  & 607 $\pm$ 123 \\
$0.05$ & 480 $\pm$ 78    &  10$\pm$ 1  \\
$0.2$  &  873 $\pm$ 222  & 0.69 $\pm$ 0.5 \\
$1$  &  2563 $\pm$ 634 &  3 $\pm$ 0.63  \\
$5$  &  2458 $\pm$ 356 &  191 $\pm$ 40  \\
$10$    &  716 $\pm$  62  &  83 $\pm$ 8  \\
\hline
\end{tabular}
\caption{\textbf{Values and [\Ca]-dependence of \IP3R2 inter-modal rates.}  Inferred \IP3R2 inter-modal transition rates \(q_{\text{PD}}\) (Park $\to$ Drive) and \(q_{\text{DP}}\) (Drive $\to$ Park) at different \Ca concentrations, 10 µM \IP3, 5 mM ATP. These parameter values are shown in Fig.~\ref{fig:dcprogs_inter_mode_single}.}
\label{tab:inter-modes rates}
\end{table}

\subsubsection*{Fitting the \Ca-dependence of the inter-modal transitions}
To facilitate the implementation of \IP3R models, we then fitted the \Ca-dependence of the inter-modal kinetic parameters \(q_{\text{PD}}\) and \(q_{\text{DP}}\) for a fixed \IP3 concentration (10 $\mu M)$.\\
For the sake of simplification, we assumed from Fig.~\ref{fig:dcprogs_inter_mode_single} and the level of noise in our inference process that is already apparent from Fig.~\ref{fig:M1_results_posteriors} and ~\ref{fig:M2_results_posteriors}, that the kinetic rate for the Park to Drive transition, \(q_{\text{PD}}\), is \Ca-independent. Thus, we fixed its value to its average over the measurements displayed in table~\ref{tab:inter-modes rates}: 1490$\pm$920 s$^{-1}$.\\
To fit the \Ca-dependence of the Drive to Park transition rate, \(q_{\text{DP}}\), we adopted the functional form proposed by Cao \textit{et al.}~\cite{cao_deterministic_2014}:
\begin{equation}
\label{eq:qDP}
q_{\text{DP}}(c) = a + V\left[1 - m(c)\,h(c)\right]
\end{equation}
where $c$ denotes the cytosolic \Ca concentration and $m(c)$ and $h(c)$ are gating variables defined as:
\begin{align}
m(c) &= \frac{c^{3}}{c^{3} + k^{3}}, &
h(c) &= \frac{p^{2}}{c^{2} + p^{2}}, \label{eq:mh}
\end{align}

%\begin{align}
%m_{24}(c) &= \frac{c^{3}}{c^{3} + k_{24}^{3}}, &
%h_{24}(c) &= \frac{k_{-24}^{2}}{c^{2} + k_{-24}^{2}}, \label{eq:m24h24}\\
%m_{42}(c) &= \frac{c^{3}}{c^{3} + k_{42}^{3}}, &
%h_{42}(c) &= \frac{k_{-42}^{3}}{c^{3} + k_{-42}^{3}}. %\label{eq:m42h42}
%\end{align}

This expression was fitted to the inferred values of \(q_{\text{DP}}\) 
shown in Fig.~\ref{fig:dcprogs_inter_mode_single}, yielding (for 10~µM 
IP$_3$ and 5~mM ATP):
\begin{align*}
    a &= 1.13~\mathrm{s^{-1}}, \\
    V &= 1255~\mathrm{s^{-1}}, \\
    k &= 9.62~\mathrm{nM}, \\
    p &= 3.70 \times 10^{4}~\mathrm{nM}.
\end{align*}

\begin{figure}
    \centering
    \includegraphics[width=0.7\linewidth]{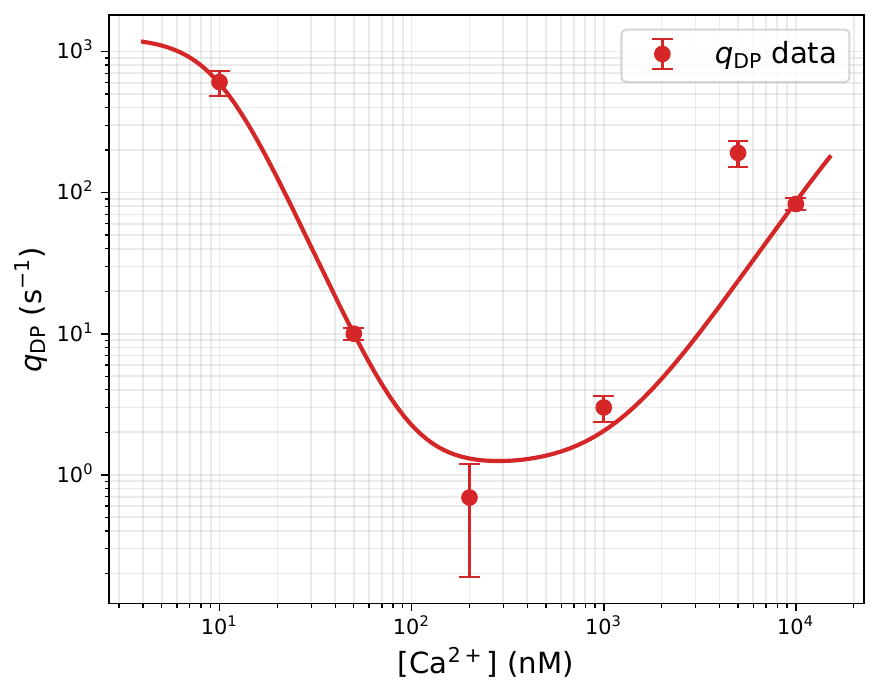}
    \caption{ \textbf{Calcium dependence of the Drive-to-Park transition rate 
    \(q_{\text{DP}}\).}
    Inferred values of the Drive-to-Park transition rate \(q_{\text{DP}}\) (red circles, mean $\pm$ s.d.) as a function of cytosolic \Ca concentration  obtained from \IP3R2 single-channel recordings at 10~µM IP$_3$ and 5~mM ATP. The solid red line shows 
    the fit of Eq.~\ref{eq:qDP}. }
    \label{fig:fit_qdp}
\end{figure}

The resulting fit is depicted in Figure~\ref{fig:fit_qdp}.

\section*{Discussion}

In this work, we revisited modal gating of IP\textsubscript{3}R2 channels using a hierarchical approach \cite{siekmann_modelling_2016} that separates intra-modal kinetics from inter-modal transitions while explicitly correcting for missed events. This strategy bridges two complementary perspectives: the continuous-time likelihood methods emphasized by Epstein et al. \cite{epstein_bayesian_2016}, which highlight the importance of accounting for undetected opening events, and the compact modal-gating framework introduced by Siekmann et al \cite{siekmann_modelling_2016}.
A first outcome of our analysis is that correcting for missed events at the intra-modal level influences the inference of fast rates. Our results support Epstein's conclusion that ignoring missed events can bias the estimation of rapid transitions, even for relatively simple models.\\

In previous studies of \IP3R modal gating modeling, to our knowledge, the exact correction of missed events was not taken into account \cite{siekmann_kinetic_2012, ullah_mode_2016}. In the absence of such correction, Bayesian model selection attributed different transition models for the Park and Drive modes: the Park mode was attributed a simple 2-state model (topology (a) in Fig.~\ref{fig:1}) whereas a more complex 4-state model was selected for the Drive mode (topology (c) in Fig.~\ref{fig:1})~\cite{siekmann_kinetic_2012,siekmann_mcmc_2012}. This suggests, in terms of protein structure, that the transition from Park to Drive would result from the emergence of two new closed states, which is not straightforward.
We show here that taking into account missed opening events clarifies this point. Our \texttt{DCPROGS}-based approach indeed suggests that when missed events are accounted for, the transition between the Park and Drive modes does not involve a change of protein structure between Park and Drive modes. For both modes, the same topology is chosen by Bayesian model selection: a three-state model with two closed state and one open state (model (b) in Fig.~\ref{fig:1}). As a result, the only change between Park and Drive modes in the final \IP3R2 model we propose is a change in the values of only two kinetic rates: the rates that govern the transitions between the two closed states of the topology. In Drive mode, these rates stabilize the closed state that is directly connected to the open state, whereas they favor the second closed state, furthest from the open state, in the Park mode. As a result, the dwell time of the closed state directly connected to the open state is much larger in Drive than in Park mode, which strongly amplifies the frequency of transitions to the open state. We believe that this prediction is more plausible from a mechanistic and molecular perspective since it does not entail a complex change of protein structure between modes.\\
In addition, taking into account missed events also simplifies the \Ca-dependence of the inter-modal transitions. In the original Park and Drive model, both the Park to Drive and the Drive to Park transitions were predicted to depend on \Ca concentration~\cite{siekmann_kinetic_2012}. Here also, our study shows that accounting for missed opening events simplifies this picture: the kinetic rate for the transition from Drive to Park remains \Ca-dependent, but the Park to Drive kinetic rate does not appear to exhibit a strong dependence on [\Ca]. Intermediate \Ca concentrations (50 nM to 10 $\mu$M) are found to strongly depress the Drive to Park transition, compared to small(< 50 nM) or large (supra-micromolar) concentrations. As a result, the \IP3R undergoes frequent transitions to the Park mode only for large or small [\Ca], and remains essentially in the Drive mode at a peak around 200 nM. This mechanism finely reproduces, at the single-molecule level, the characteristic bell-shaped dependence of \IP3R activity on \Ca concentration observed in population experiments~\cite{bezprozvanny_bell-shaped_1991}. Moreover, this mechanism aligns favorably with the idea that modal gating reflects slow conformational rearrangements whose relative stability depends on \Ca concentration.\\

An important practical challenge of the \texttt{DCPROGS} approach is that it is based on a burst analysis (see Methods). In particular, a major constraint is that the method requires a burst to begin with an open state. In our case, however, the intra-modal segments contain many transitions and, more importantly, missed events are already integrated directly into the likelihood. This reduces the impact of this constraint.
However, our analysis is limited by other practical issues. First, we hypothesized that the experimental dataset we used corresponds to the activity of exactly one \IP3R channel in the patch. Violation of this hypothesis would strongly influence the overall likelihood structure. An improvement of our method, kept for future work, would therefore consist in inferring the number of \IP3R channel molecules in each patch clamp recording \cite{horn_estimating_1991, colquhoun_stochastic_1990} before inferring the model topology itself. Secondly, an important limitation of our study is that the admissible intra-modal model topologies were limited to the four models shown in Fig.~\ref{fig:1}. Recent studies have proposed to use deep learning models to infer the correct topology of a Markov model among a large number of candidate topologies, chosen impartially \cite{oikonomou_deep_2024}. Incorporating these methods into our approach would strongly improve its robustness, albeit at the price of a strong inflation in the necessary computing resources.
Finally, the accuracy of our inference of the intra-modal kinetic parameters could be improved. The posterior distributions (Fig.~\ref{fig:M1_results_posteriors}A-B and ~\ref{fig:M2_results_posteriors}A-B) attest to the quality of our inferences. However, the strong variability of the inference for kinetic parameters that are hypothesized to be \Ca-independent suggests that the accuracy of our inference methods could be improved. Beyond methodological improvements, a practical way to increase accuracy would be to benefit from larger datasets, in particular patch-clamp recordings of longer duration. Lastly, we note that a similar implementation could be a Hidden Markov model relying on a unique 3-state topology, with the values of the transition rates between its two closed states determined by a latent Markov Model that sets the probability to observe Drive or Park. Controlling the parameters of this latent model with the \Ca concentration would then allow reproducing the \Ca-dependence of Fig.~\ref{fig:dcprogs_inter_mode_single}. Future work is needed to determine how to calibrate such a hidden Markov model and whether it provides a better description of \IP3R channel dynamics than the model proposed here. \\  

The final output of our work is a new Bayesian model for \IP3R channels with multi-modal gating. This model, presented in Fig.~\ref{fig:siekmann_comp_single}, table~\ref{tab:siekmann_parameters}, eqs.~\eqref{eq:qDP} and ~\eqref{eq:mh}, proposes that both the Park and Drive modes consist of the 3-state topology of Fig.~\ref{fig:1}b with two closed states and one open state. According to the model, all the intra-modal transition rates in both modes are \Ca-independent, the only difference between the two modes lying in the transition rates between the two closed states, that specify the Park or Drive mode. In the implementation we propose here, inter-modal transitions connect the two closed states that are directly connected to the open one (Fig.~\ref{fig:siekmann_comp_single}), with only the Drive to Park rate that is \Ca-dependent (eqs.~\eqref{eq:qDP} and ~\eqref{eq:mh}).

\section*{Author Contributions}
S.B. performed the research and contributed analytic tools; S.B., A.D., H.B. designed the research; and S.B., A.D., H.B. wrote the article.

\section*{Acknowledgments}
The authors thank Dr.\ Siekmann for kindly providing the experimental IP\textsubscript{3}R datasets, and Drs.\ Wagner and Yule for granting permission to use their patch-clamp recordings. We also thank Dr. Jan-Michael Rye for maintaining a fork of the DCPROGS/HJCFIT software and implementing minor bug fixes.

The authors acknowledge financial support by the French Agence Nationale de la Recherche (ANR), project SecNet under reference \href{https://anr.fr/Project-ANR-22-CE16-0034}{ANR-22-CE16-0034}.

\section*{Declaration of Interests}
The authors declare no competing interests.
\bibliographystyle{abbrv}
\bibliography{ip3R2.bib}

\end{document}